\documentclass[12pt,
showpacs,amsmath,amssymb]{revtex4}
\def\nc{\newcommand}
\nc{\be}{\begin{equation}}
\nc{\ee}{\end{equation}}
\def\ba{\begin{array}{ll}}
\def\ea{\end{array}}
\nc{\bea}{\begin{eqnarray}}
\nc{\eea}{\end{eqnarray}}
\def\nn{\nonumber}

\nc{\bl}{\mathbf l}
\nc{\bn}{\mathbf n}
\nc{\bm}{\mathbf m}
\nc{\cL}{\mathcal{L}}
\def\dL{\mathcal{L}^{\dagger}}

\nc{\cD}{\mathcal{D}}
\def\sqd{\sqrt{2}}

\def\p{\partial}
\def\pr{\p_r}
\def\pv{\p_v}
\def\pta{\p_{\theta}}
\def\pvi{\p_{\varphi}}
\def\spr{\frac{\p}{\p r_*}}

\nc{\pd}[1]{\frac{\p}{\p{#1}}}
\nc{\pdo}[1]{\frac{\p^2}{\p{#1}^2}}
\nc{\pdt}[2]{\frac{\p^2}{\p{#1}\p{#2}}}
\def\spdr{\frac{\p^2}{\p r_*^2}}
\def\spdvr{\frac{\p^2}{\p r_* \p v_*}}
\def\spdra{\frac{\p^2}{\p r_* \p \theta_*}}

\def\spdrp{\frac{\p^2}{\p r_* \p \varphi_*}}
\def\sta{\sin\theta}
\def\cta{\cos\theta}
\def\coa{\cot\theta}
\def\sda{\sin^2\theta}

\begin{document}
\baselineskip 18pt
\begin{flushright}
USTC-ICTS 03-9 \\
gr-qc/0306044 \\
\end{flushright}
\title{\bf Hawking Radiation of an Arbitrarily Accelerating Kinnersley Black Hole:
Spin-Acceleration Coupling Effect \footnote{Supported partially by China Postdoctoral
Science Foundation and K.C. Wong Education Foundation, Hong Kong.}}
\author{WU Shuang-Qing 
$^*$, and YAN Mu-Lin 
$^{\dagger}$}
\affiliation{\it\footnotesize Interdisciplinary Center for Theoretical Study,
$\&$ Department of Modern Physics, \\
\it\footnotesize University of Science and Technology, Hefei 230026, People's
Republic of China \\
\rm $^*$E-mail: sqwu@ustc.edu.cn ~~$^{\dagger}$E-mail: mlyan@ustc.edu.cn}
\date{revised 17/7, 2003}

\begin{abstract}
{\em The Hawking radiation of Weyl neutrinos in an arbitrarily accelerating Kinnersley
black hole is investigated by using a method of the generalized tortoise coordinate
transformation. Both the location and temperature of the event horizon depend on the
time and on the angles. They coincide with previous results, but the thermal radiation
spectrum of massless spinor particles displays a kind of spin-acceleration coupling
effect.}
\end{abstract}

\maketitle

In recent ten years, the study of Hawking radiation of some non-stationary black holes,
especially with a method of the generalized tortoise coordinate transformation (GTCT),
has attracted more attention and achieved much progress.$^{1,2}$ The tortoise coordinate
transformation method initially proposed by Damour and Ruffini$^3$ to deal with the Hawking
effect of scalar fields or Dirac particles in some static spherically symmetric black holes
and stationary axisymmetric space-times,$^4$ has been generalized by Zhao \textit{et al}$^5$
to the non-static and non-stationary cases and successfully applied to investigate
quantum thermal effect of scalar particles in various non-static spherically symmetric
space-times and non-stationary black holes.$^1$

As far as spinor field case is concerned, the GTCT method has no difficulty in discussing
Hawking evaporation of Dirac particles in some non-static spherically symmetric black holes
also,$^1$ but it is not so easy to use this method to study quantum thermal effect of fermions
in the non-static and non-spherically symmetric black holes and in the non-stationary black
holes. The difficulty lies in the non-separability of Chandrasekhar-Dirac equation$^6$ in the
most general space-times. Until quite recently this difficulty within the GTCT formalism has
been overcome by one of us$^2$ via simultaneously treating the first-order and second-order
Dirac equations so that one can recast each second-order equation for four spinor components
into a standard wave equation near the event horizon. With this extended GTCT method, we
first succeed in dealing with the Hawking radiation of Dirac particles in a variable-mass
Kerr(-Newman) black hole and observe a new quantum effect called as spin-rotation coupling
effect appearing in the thermal radiation spectrum of fermions.$^7$ This kind of quantum
effect is absent from the bosonic radiation spectrum of scalar particles. Then we reexamine
the Hawking evaporation of higher spin particles in the non-static spherically symmetric
black holes and show that the Hawking radiation is asymmetric amongst different components
of higher spin fields.$^8$ Finally we extend this treatment to cope with the Hawking effect
of spinor particles in the Kinnersley black hole.$^{9,10}$ Our results show that a
spin-acceleration coupling effect displays in the thermal radiation spectrum of Dirac
particles in the arbitrarily accelerating Kinnersley black hole$^{10}$ while it does not
present in that of fermions in the rectilinearly accelerating Kinnersley black hole.$^9$
In all cases we demonstrate that the event horizon equation and Hawking temperature coincide
with previously published results.

In this letter, we reinvestigate quantum thermal effect of an arbitrarily accelerating
Kinnersley black hole,$^{10}$ but adopt another null tetrad system different from that of
Ref. [10]. For simplicity we only consider the case of massless spinor particles namely Weyl
neutrinos because the Hawking radiation has of no relation to the mass of particles.$^2$ We
show that the location of the event horizon and Hawking temperature of the Kinnersley black
hole are independent of the choice of a concrete null tetrad system, they coincide with those
previously obtained results, however the expression of spin-acceleration effect is much
simpler than that obtained in Ref. [10].

The Kinnersley metric,$^{11}$ generally called as the ``photon rocket'' solution, is
interpreted as the external gravitational field of an arbitrarily accelerating mass. In
the advanced Eddington-Finkelstein coordinate system, the line element of Kinnersley's
black hole can be rewritten as
\be \label{eq1}
ds^2 = 2dv(F dv -dr) -r^2\big[(d\theta +f dv)^2 +\sda(d\varphi +g dv)^2\big] \, ,
\ee
where $2F = 1 -2M(v)/r -2ar\cta$, $f = b\sin\varphi +c\cos\varphi -a\sta$, and $g =
(b\cos\varphi -c\sin\varphi)\coa$. The arbitrary function $M(v)$ describes the change
in the mass of the source as a function of the advanced time; $a = a(v)$, $b = b(v)$
and $c = c(v)$ are acceleration parameters: $a$ is the magnitude of acceleration, $b$
and $c$ are the rates of change of its direction.

Now we choose such a complex null-tetrad system $\{\bl, \bn, \bm, \overline{\bm}\}$
that it satisfies the pseudo-orthogonal condition $\bl \cdot \bn = -\bm \cdot
\overline{\bm} = 1$ in which the
directional derivatives are
\bea
&& D = -\pr \, , \qquad\qquad\quad \Delta = \pv +F\pr -f\pta -g\pvi \, , \nn \\
&& \delta = \frac{1}{\sqd r}\big(\pta +\frac{i}{\sta}\pvi\big) \, , \quad
\overline{\delta} = \frac{1}{\sqd r}\big(\pta -\frac{i}{\sta}\pvi\big) \, .
\eea
It is not difficult to determine the non-vanishing Newman-Penrose complex spin
coefficients$^{12}$ in the above null-tetrad as follows [Here and hereafter, we
denote $F_{,r} = dF/dr$, etc.]
\bea
&& \rho = \frac{1}{r} \, , \quad \mu = \frac{F}{r} -f_{,\theta} \, , \quad
\gamma = \frac{1}{2}\big(-F_{,r} +\frac{ig}{\cta}\big) \, , \nn \\
&& \beta = -\alpha = \frac{\coa}{2\sqd r} \, , \quad
\nu = \frac{F_{,\theta}}{\sqd r} = \frac{a\sta}{\sqd} \, .
\eea

In the case of the Kinnersley black hole, $\kappa = \lambda = \sigma = \epsilon =
\tau = \pi = 0$ and $\psi_3 = 0$, but $\nu \not= 0$, thus it is of Petrov type D.
Following Teukolsky's black hole perturbation theory,$^6$ one can start from the
Weyl neutrino equation
\bea \label{eq4}
&& (D +\epsilon -\rho)\eta_1 -(\overline{\delta} +\pi -\alpha)\eta_0 = 0 \, , \nn \\
&& (\delta +\beta -\tau)\eta_1 -(\Delta +\mu -\gamma)\eta_0 = 0 \,
\eea
to derive its corresponding second-order perturbed wave equation
\bea \label{eq5}
&& \big[(D +\epsilon^* -\rho -\rho^*)(\Delta +\mu -\gamma) -(\delta -\alpha^*
-\tau +\pi^*)(\overline{\delta} +\pi -\alpha)\big]\eta_0 = 0 \, , \nn \\
&& \big[(\Delta -\gamma^* +\mu +\mu^*)(D +\epsilon -\rho)
-(\overline{\delta} +\beta^* +\pi -\tau^*)(\delta +\beta -\tau)\big]\eta_1 \nn \\
&&\qquad\qquad\qquad\qquad\qquad\qquad
\qquad= \big[\nu D +(D +2\epsilon +\epsilon^* -\rho^*)\nu -\psi_3\big]\eta_0 \, .
\eea

Write these wave equations (\ref{eq4}) and (\ref{eq5}) in the above null-tetrad,
one obtains their concrete expressions in the Kinnersley black hole
\bea \label{eq6}
(\pr +\frac{1}{r})\eta_1 +\frac{1}{\sqd r}\cL_{1/2}\eta_0 = 0 \, , \nn &&\\
\big(\mathcal{G} -f_{,\theta} -\frac{ig}{2\cta}\big)\eta_0
-\frac{1}{\sqd r}\dL_{1/2}\eta_1 = 0 \, &&
\eea
and
\bea \label{eq7}
&& \big[2r^2(\pr +2/r)(\mathcal{G} -f_{,\theta}-\frac{ig}{2\cta})
+\dL_{1/2}\cL_{1/2}\big]\eta_0 =  0 \, , \nn \\
&& \big[2r^2(\mathcal{G} +F/r -2f_{,\theta} +\frac{ig}{2\cta})(\pr +1/r)
+\cL_{1/2}\dL_{1/2}\big]\eta_1 \nn \\
&&\qquad\qquad\qquad\qquad\qquad\qquad = \sqd r^2a\sta(\pr +1/r)\eta_0  \, ,
\eea
in which the operators $\mathcal{G}$, $\cL_n$ and $\dL_n$ are defined as
\bea
&&\qquad\qquad \mathcal{G} = \pv +F\pr -f\pta -g\pvi +F/r +F_{,r}/2 \, , \nn \\
&& \cL_n = \pta +n\coa -\frac{i}{\sta}\pvi \, , \quad
\dL_n = \pta +n\coa +\frac{i}{\sta}\pvi \, . \nn
\eea

After making substitutions $\chi_1 = \sqd r\eta_1$ and $\chi_0 = \eta_0$, Eqs.
(\ref{eq6}) and (\ref{eq7}) become
\bea \label{eq8}
\pr\chi_1 +\cL_{1/2}\chi_0 =  0 \, , \nn &&\\
2r^2(\mathcal{G} -f_{,\theta} -\frac{ig}{2\cta}\big)\chi_0
-\dL_{1/2}\chi_1 = 0 \, &&
\eea
and
\bea \label{eq9}
&& \big[2r^2(\pr +2/r)(\mathcal{G} -f_{,\theta}-\frac{ig}{2\cta})
+\dL_{1/2}\cL_{1/2}\big]\chi_0 =  0 \, , \nn \\
&& \big[2r^2(\mathcal{G} -2f_{,\theta} +\frac{ig}{2\cta})\pr
+\cL_{1/2}\dL_{1/2}\big]\chi_1 = 2r^3a\sta(\pr +1/r)\chi_0  \, .
\eea

As there exists no Killing vector in an arbitrarily accelerating Kinnersley black
hole, one can define the most general form of GTCT as did in Ref. [1,2,10]
\be \label{eq10}
r_* = r +\frac{1}{2\kappa}\ln \big(r -r_H\big) \, , \quad
v_* = v -v_0 \, , \quad \theta_* = \theta -\theta_0 \, , \quad
\varphi_* = \varphi -\varphi_0 \, ,
\ee
where $r_H = r_H(v,\theta,\varphi)$ is the location of the event horizon, $\kappa$
is an adjustable parameter and is unchanged under tortoise transformation. All
parameters $v_0, \theta_0$ and $\varphi_0$ are arbitrary constants characterizing
the initial state of the hole.

Applying the GTCT (\ref{eq10}) to the first-order equation (\ref{eq8}) and taking
the $r \rightarrow r_H(v_0,\theta_0,\varphi_0)$, $v \rightarrow v_0$, $\theta
\rightarrow \theta_0$ and $\varphi \rightarrow \varphi_0$ limits, near the
event horizon Eq. (\ref{eq8}) reduces to
\bea \label{eq11}
\big(r_{H,\theta} +\frac{i}{\sta_0}r_{H,\varphi}\big)\spr\chi_1 +2r_H^2\big(F
-r_{H,v} +fr_{H,\theta} +gr_{H,\varphi}\big)\spr \chi_0 = 0 \, , \nn &&\\
-\spr\chi_1 +\big(r_{H,\theta} -\frac{i}{\sta_0}r_{H,\varphi}\big)\spr\chi_0 = 0 \, . &&
\eea
From the vanishing of the determinant of Eq. (\ref{eq11}), one gets the following
equation that determines the location of event horizon
\be \label{eq12}
2(F -r_{H,v} +f r_{H,\theta} +2g r_{H,\varphi}) +\frac{r_{H,\theta}^2}{r_H^2}
+\frac{r_{H,\varphi}^2}{r_H^2\sda_0} = 0 \, ,
\ee
which is just the same one derived in Ref. [10]. It should be noted that the
location $r_H$ we obtained above is just a local point on the section of null
surface of the event horizon. This is enough because the GTCT method is a local
analysis one. One only needs to calculate the value of the event horizon $r_H$
at the specific moment $v_0$ and angles $(\theta_0, \varphi_0)$. It corresponds
to make a analysis locally at every point on the event horizon of a non-stationary
black hole (see the last one of Ref. [9] for a complete discussion about this point).

Expand and write out the second-order equation (\ref{eq9}) obviously, we have
\bea \label{eq13}
&& \Big\{2r^2\big[\p_{vr}^2 +F\pr^2 -f\p_{r\theta}^2 -g\p_{r\varphi}^2\big]
+(3r^2F_{,r} +6rF -2r^2f_{,\theta} -r^2\frac{ig}{\cta})\pr \nn \\
&&\quad +4r(\pv -f\pta -g\pvi) +\pta^2 +\frac{1}{\sda}\pvi^2
+\coa\pta +\frac{i\cta}{\sda}\pvi \nn \\
&&\quad -\frac{1}{4\sda} -\frac{1}{4} +r^2F_{,rr} +4rF_{,r} +2F
-4rf_{,\theta} -2r\frac{ig}{\cta})\Big\}\chi_0 = 0 \, \quad
\eea
and
\bea \label{eq14}
&& \Big\{2r^2\big[\p_{vr}^2 +F\pr^2 -f\p_{r\theta}^2 -g\p_{r\varphi}^2\big]
+(r^2F_{,r} +2rF -4r^2f_{,\theta} +2r^2\frac{ig}{\cta})\pr +\pta^2 \nn \\
&&\quad +\frac{1}{\sda}\pvi^2 +\coa\pta -\frac{i\cta}{\sda}\pvi
-\frac{1}{4\sda} -\frac{1}{4}\Big\}\chi_1 = 2r^3a\sta(\pr +1/r)\chi_0 \, .
\eea
In addition, the Klein-Gordon equation of scalar particles with mass $\mu_0$:
$(\Box +\mu_0^2)\Phi = 0$ in the Kinnersley space-time can be written explicitly
as
\bea \label{eq15}
&& \Big\{2r^2\big[\p_{vr}^2 +F\pr^2 -f\p_{r\theta}^2 -g\p_{r\varphi}^2\big]
+(2r^2F_{,r} +4rF -2r^2f_{,\theta})\pr \nn \\
&&\quad +2r(\pv -f\pta -g\pvi) +\pta^2 +\frac{1}{\sda}\pvi^2
+\coa\pta -\mu_0^2r^2\Big\}\Phi = 0 \, . \qquad
\eea

Now let us consider the asymptotic behaviors of the second-order wave equation
near the event horizon. Under the transformation (\ref{eq10}), the limiting
forms of Eqs. (\ref{eq13}-\ref{eq15}) are
\bea \label{eq16}
&&\tilde{\mathcal{K}} \chi_0
+\Big[-\widetilde{A} +\frac{3F}{r_H} +\frac{3F_{,r}}{2}
-f_{,\theta} -\frac{ig}{2\cta_0} -\frac{i\cta_0}{2r_H^2\sda_0}
r_{H,\varphi}
-\frac{\coa_0}{2r_H^2}r_{H,\theta} \nn \\
&&\qquad\qquad +\frac{2}{r_H}\big(fr_{H,\theta} +gr_{H,\varphi}
-r_{H,v}\big) -\big(\frac{r_{H,\theta\theta}}{2r_H^2}
+\frac{r_{H,\varphi\varphi}}{2r_H^2\sda_0}\big)\Big]\spr\chi_0 = 0 \,
\eea
and
\bea \label{eq17}
&& \tilde{\mathcal{K}} \chi_1 +\Big[-\widetilde{A} +\frac{F}{r_H} +\frac{F_{,r}}{2}
-2f_{,\theta} +\frac{ig}{2\cta_0} +\frac{i\cta_0}{2r_H^2\sda_0}r_{H,\varphi}
-\frac{\coa_0}{2r_H^2}r_{H,\theta} \nn \\
&&\qquad\qquad\qquad\qquad\qquad -\big(\frac{r_{H,\theta\theta}}{2r_H^2}
+\frac{r_{H,\varphi\varphi}}{2r_H^2\sda_0}\big)\Big]\spr\chi_1 =
r_Ha\sta_0 \spr\chi_0 \,
\eea
as well as
\bea \label{eq18}
&& \tilde{\mathcal{K}} \Phi +\Big[-\widetilde{A} +\frac{2F}{r_H} +F_{,r}
-f_{,\theta} +\frac{1}{r_H}\big(fr_{H,\theta} +gr_{H,\varphi} -r_{H,v}\big)\nn \\
&&\qquad\qquad -\frac{\coa_0}{2r_H^2}r_{H,\theta}-\big(\frac{r_{H,\theta\theta}}{2r_H^2}
+\frac{r_{H,\varphi\varphi}}{2r_H^2\sda_0}\big)\Big]\spr\Phi = 0 \, ,
\eea
where the operator $\tilde{\mathcal{K}}$ represents the second derivative term
\bea
&& \tilde{\mathcal{K}} = \big[\frac{\widetilde{A}}{2\kappa} +2F
-r_{H,v} +f r_{H,\theta} +g r_{H,\varphi}\big]\spdr +\spdvr \nn \\
&&\qquad -\big(f +\frac{r_{H,\theta}}{r_H^2}\big)\spdra
-\big(g +\frac{r_{H,\varphi}}{r_H^2\sda_0}\big)\spdrp \, , \nn
\eea
and the coefficient $\widetilde{A}$ is an infinite limit of $0/0$-type
\bea
&& \widetilde{A} = \lim_{r\to r_H, v\to v_0 \atop {\theta\to \theta_0,
\varphi\to \varphi_0}} \frac{F -r_{H,v} +f r_{H,\theta} +gr_{H,\varphi}
+r_{H,\theta}^2/(2r^2) +r_{H,\varphi}^2/(2r^2\sda)}{r -r_H} \nn \\
&&\quad = F_{,r} -\big(\frac{r_{H,\theta}^2}{r_H^3}
+\frac{r_{H,\varphi}^2}{r_H^3\sda_0}\big) \, .
\eea

Let the coefficient in front of the second-order derivative term $\spdr$
be $1/2$, namely
$$\frac{\widetilde{A}}{2\kappa} +2F -r_{H,v} +f r_{H,\theta}
+g r_{H,\varphi} \equiv \frac{1}{2} \, ,$$
then the adjustable temperature parameter $\kappa$ can be determined as
\be \label{eq20}
\kappa = \frac{r_H^2F_{,r} -r_H^{-1}(r_{H,\theta}^2
+r_{H,\varphi}^2/\sda_0)}{r_H^2(1 -2F) +(r_{H,\theta}^2
+r_{H,\varphi}^2/\sda_0)} \, ,
\ee
and it coincides with the Hawking temperature derived in Ref. [10].

With such a parameter adjustment, one can recast Eqs. (\ref{eq16}-\ref{eq18})
into an united standard wave equation near the event horizon
\be \label{eq21}
\big[\spdr +2\spdvr -2C_3\spdra -2\Omega\spdrp +2(C_2 +iC_1)\spr\big]\Psi = 0 \, ,
\ee
where
$$\Omega = g +\frac{r_{H,\varphi}}{r_H^2\sda_0} \, , \qquad
C_3 = f +\frac{r_{H,\theta}}{r_H^2} \, .$$
In the Weyl neutrino case, we have
\bea
&& C_2 +iC_1 = \frac{F}{r_H} +\frac{F_{,r}}{2} -f_{,\theta}
-\frac{ig}{2\cta_0} -\frac{i\cta_0}{2r_H^2\sda_0}r_{H,\varphi} \nn \\
&&\qquad\qquad\quad -\frac{\coa_0}{2r_H^2}r_{H,\theta}
-\frac{1}{2r_H^2}\big(r_{H,\theta\theta}
+\frac{r_{H,\varphi\varphi}}{\sda_0}\big) \, ,
\qquad\qquad\qquad\qquad (\Psi = \chi_0) \, , \nn \\
&&2(C_2 +iC_1) =
\frac{F}{r_H} -\frac{F_{,r}}{2} -2f_{,\theta}
+\frac{ig}{\cta_0} +\frac{i\cta_0}{2r_H^2\sda_0}r_{H,\varphi} \nn \\
&&\qquad\qquad\qquad -\frac{\coa_0}{2r_H^2}r_{H,\theta}
-\frac{1}{2r_H^2}\big(r_{H,\theta\theta}
+\frac{r_{H,\varphi\varphi}}{\sda_0}\big) \nn \\
&&\qquad\qquad\qquad +\frac{1}{r_H^3}\big(r_{H,\theta}^2
+\frac{r_{H,\varphi}^2}{\sda_0}\big) -ar_H\sta_0\big(r_{H,\theta}
-\frac{ir_{H,\varphi}}{\sta_0}\big)^{-1} \, ,
\quad (\Psi = \chi_1) \, ; \nn
\eea
whereas for scalar fields ($\Psi = \Phi$), we have $C_1 = 0$ and
\bea
&& C_2 = \frac{F}{r_H} -f_{,\theta} -\frac{r_{H,\theta}}{2r_H^2}\coa_0
-\frac{1}{2r_H^2}\big(r_{H,\theta\theta} +\frac{r_{H,\varphi\varphi}}{\sda_0}\big)
+\frac{1}{2r_H^3}\big(r_{H,\theta}^2 +\frac{r_{H,\varphi}^2}{\sda_0}\big) \, . \nn
\eea

Following the method of Damour-Ruffini-Sannan's,$^3$ the thermal radiation spectra
of scalar particles and Weyl neutrinos can be obtained from the standard wave equation
(\ref{eq21}) near the event horizon$^{10}$
\be \label{eq22}
\langle \mathcal{N}(\omega) \rangle \simeq \frac{1}{e^{(\omega
+m\Omega -C_1)/T_H}\pm 1} \, , \qquad T_H = \frac{\kappa}{2\pi} \, ,
\ee
in which the coefficient $C_1 = 0$ (for $\Psi = \Phi$); while for $\Psi =
\chi_0, \chi_1$, it reads
\bea \label{eq23}
C_1 &=& -\frac{g}{2\cta_0} -\frac{r_{H,\varphi}\cta_0}{2r_H^2\sda_0}  \, ,
\qquad\quad (\Psi = \chi_0) \, , \nn \\
C_1 &=& \frac{g}{2\cta_0} +\frac{r_{H,\varphi}\cta_0}{2r_H^2\sda_0} \nn \\
&& -ar_Hr_{H,\varphi}\big(r_{H,\theta}^2 +\frac{r_{H,\varphi}^2}{\sda_0}\big)^{-1} \, ,
\quad (\Psi = \chi_1) \, .
\eea

It is easy to see that the last term in the right hand side of the second one
of Eq. (\ref{eq23}) is proportional to the spin coefficient $\nu$. If we neglect
this term because it is not invariant under the null tetrad transformation, then
the ``spin-dependent'' term can be written as
\bea \label{eq24}
\omega_s &\simeq& s\big(\frac{r_{H,\varphi}\cta_0}{r_H^2\sda_0}
+\frac{g}{\cta_0}\big) \, , \\
&& (s = -1/2, 1/2 \, , ~\Psi = \chi_0\, , \chi_1 \, ; \quad s = 0
\, , ~\Psi = \Phi) \, . \nn
\eea
The term $\omega_s$ reflects that there exists a kind of spin-acceleration coupling
effect in the arbitrarily accelerating Kinnersley black hole.$^{2,10}$

In this paper, we have studied the quantum thermal effect of an arbitrarily
accelerating Kinnersley black hole. Eqs. (\ref{eq12}) and (\ref{eq20}) give
the location and temperature of the event horizon of the Kinnersley black hole,
which depend not only on the advanced time $v$ but also on the angles $\theta$
and $\varphi$. Eq. (\ref{eq22}) shows the thermal radiation spectra of scalar
particles and Weyl neutrinos in an arbitrarily rectilinearly accelerating
Kinnersley black hole. It manifests that there is a new spin-acceleration coupling
effect in the arbitrarily accelerating Kinnersley black hole while it disappears
in the rectilinearly accelerating Kinnersley black hole. This effect does not
present in the thermal radiation spectrum of scalar particles also.

It should be noted that a null tetrad system different from that of Ref. [10] has
been chosen here. The difference between them is a Lorentz rotation of type I,
however both leads to the same results (the event horizon equation, the Hawking
temperature and the thermal radiation spectrum). The expression of the coefficients
$C_1$ and $C_2$ here has a slight difference from that of Ref. [10], it is much simpler
than the latter. Besides, both of them have a similar structure, especially the
coefficient $C_1$ is the same for the $s = -1/2$ component. However the coefficients $C_1$
and $C_2$ for the $s = 1/2$ component are different in different null tetrad systems,
in which there is a term proportional to the spin coefficient $\nu$. The difference
probably relates to the shift of the $s = 1/2$ spinor component under the Lorentz
rotation of type I. Thus the term proportional to the spin coefficient $\nu$
can be thought as a gauge dependent term and should be ignored. Though the results
have a little difference, the thermal spectrum of spinor particles still reflects
qualitatively that there exists a spin-acceleration coupling effect in the arbitrarily
accelerating Kinnersley black hole.

In summary, another null tetrad system distinguished from Ref. [10] is chosen to
investigate the Hawking effect of the Kinnersley black hole. The choice of basis
is more convenient for computations here, and thus the final results are more concise
and compact. However we find that the event horizon equation, the Hawking temperature
and the thermal radiation spectrum are consistent with those derived in Ref. [10].
We think it should arrive at the same results by choosing different coordinate systems
to discuss the Hawking radiation, and conclude that all physical results should be
independent of the choice of a concrete coordinate system.

\end{document}